\documentclass[conference]{IEEEtran}

\usepackage{cite}
\usepackage[cmex10]{amsmath}
\usepackage{graphicx}
\usepackage{picinpar}
\usepackage[cmex10]{amsmath}
\usepackage{amsmath,amsfonts,amssymb}
\usepackage{array}

\usepackage{mdwmath}
\usepackage{mdwtab}
\usepackage{eqparbox}
\usepackage{subfigure}
\usepackage{algorithm}
\usepackage{algorithmic}
\usepackage{stfloats}
\usepackage{fixltx2e}
\usepackage{bm}

\begin{document}
\newtheorem{lemma}{Lemma}
\newtheorem{corol}{Corollary}
\newtheorem{theorem}{Theorem}
\newtheorem{proposition}{Proposition}
\newtheorem{definition}{Definition}
\newcommand{\e}{\begin{equation}}
\newcommand{\ee}{\end{equation}}
\newcommand{\eqn}{\begin{eqnarray}}
\newcommand{\eeqn}{\end{eqnarray}}
\renewcommand{\algorithmicrequire}{ \textbf{Input:}} 
\renewcommand{\algorithmicensure}{ \textbf{Output:}} 

\title{Compressive Massive Random Access for Massive Machine-Type Communications (mMTC)}

\author{\IEEEauthorblockN{Malong Ke\IEEEauthorrefmark{2}\IEEEauthorrefmark{1}, Zhen Gao\IEEEauthorrefmark{2}\IEEEauthorrefmark{1}, Yongpeng Wu \IEEEauthorrefmark{3}, and Xiangming Meng \IEEEauthorrefmark{4}}
\IEEEauthorblockA{\IEEEauthorrefmark{2}Advanced Research Institute of Multidisciplinary Science, Beijing Institute of Technology, Beijing 100081, China}
\IEEEauthorblockA{\IEEEauthorrefmark{1}School of Information and Electronics, Beijing Institute of Technology, Beijing 100081, China}
\IEEEauthorblockA{\IEEEauthorrefmark{3}Shanghai Jiao Tong University, Shanghai 200240, China}
\IEEEauthorblockA{\IEEEauthorrefmark{4}Huawei Technologies Co.Ltd., Shanghai 201206, China}
Email: gaozhen16@bit.edu.cn}

\maketitle

\begin{abstract}
In future wireless networks, one fundamental challenge for massive machine-type communications (mMTC) lies in the reliable support of massive connectivity with low latency.
Against this background, this paper proposes a compressive sensing (CS)-based massive random access scheme for mMTC by leveraging the inherent sporadic traffic, where both the active devices and their channels can be jointly estimated with low overhead.
Specifically, we consider devices in the uplink massive random access  adopt pseudo random pilots, which are designed under the framework of CS theory.
Meanwhile, the massive random access at the base stations (BS) can be formulated as the sparse signal recovery problem by leveraging the sparse nature of active devices.
Moreover, by exploiting the structured sparsity among different receiver antennas and subcarriers, we develop a distributed multiple measurement vector approximate message passing (DMMV-AMP) algorithm for further improved performance.
Additionally, the state evolution (SE) of the proposed DMMV-AMP algorithm is derived to predict the performance.
Simulation results demonstrate the superiority of the proposed scheme, which exhibits a good tightness with the theoretical SE.
\end{abstract}

\begin{IEEEkeywords}
Massive random access, massive machine-type communications (mMTC), compressive sensing (CS).
\end{IEEEkeywords}

\section{Introduction}
Driven by the Internet-of-Things (IoT), how to support massive machine-type communications (mMTC) enabling massive connectivity with tens of billions of machine-type devices has been challenging the current  wireless networks \cite{{5G:mMTC}}.
For mMTC, how to reliably support the access and estimate the associated channels of active devices triggered by external events dynamically performs an important role in uplink systems.

Conventional grant-based access control requires the additional control signaling and the prediction of uplink access requests for the granting of resources \cite{{Grant-based protocol 1}}.
One representative is the strongest-user collision resolution (SUCRe) scheme \cite{{Grant-based protocol 2}}, which designs a sophisticated protocol efficiently supporting devices in overloaded networks.
However, the grant-based solutions may suffer from the inefficient schedule and difficult design for the access in mMTC \cite{{Grant-based protocol 1}}.
As an alternative, grant-free access protocol without prior scheduling permission has recently attracted significant attention \cite{{Grant-free}}.
By exploiting the inherent sporadic traffic, a compressive sensing (CS)-based uplink grant-free non-orthogonal multiple access (NOMA) scheme has been proposed to further reduce the overhead \cite{{Grant-free CS}, {Grant-free CS_2}}, while only single-antenna is considered at the base station (BS).
In multi-antenna systems, a modified Bayesian compressive sensing (BCS) algorithm is proposed for NOMA \cite{{CS MIMO}}, which exploits the structured sparsity among different receiver antennas to enhance the performance.
However, the BCS algorithm may not work efficiently in mMTC. Recently, a low-complexity iterative algorithm termed approximate message passing (AMP) has been proposed for massive random access \cite{{AMP SE}, {AMP_EP}, {Bayesian}}, which can support the massive connectivity efficiently.
However, most prior work  \cite{{Grant-based protocol 1},{Grant-based protocol 2},{Grant-free},{Grant-free CS},{CS MIMO},{AMP SE}} are limited to the single-carrier systems.
Besides, the work in \cite{{AMP SE}} assumes the full knowledge of the prior distribution and noise variance, which is an impractical assumption.

In this paper, we make further research in the direction of massive random access for mMTC, where we consider the more practical frequency-selective fading channels and the work is extended to the multi-carrier multi-antenna systems.
Specifically, by exploiting the structured sparsity among different receiver antennas and subcarriers, we develop a distributed multiple measurement vector approximate message passing (DMMV-AMP) algorithm for further improved performance.
Moreover, resorting to the expectation maximization (EM) algorithm, the proposed DMMV-AMP algorithm can learn unknown hyperparameters of the prior distribution and noise variance.
We further derive the state evolution (SE) of DMMV-AMP algorithm to characterize the performance of the proposed scheme.

\section{System Model}

We consider a typical uplink mMTC system with one BS equipped with $M$ antennas and $K$ single-antenna devices, where OFDM with $N$ subcarriers is adopted to combat the time dispersive channels, and $P$ pilots are uniformly allocated across $N$ subcarriers.
For the subchannel at the $p$-th subcarrier $\left( {1 \le p \le P} \right)$, the received signal ${{\bf{y}}_{p,k}^t}{ \in \mathbb{C}^{M \times 1}}$ at the BS from the $k$-th device during the $t$-th OFDM symbol can be expressed as
\begin{equation}
\begin{array}{l}
{\bf{y}}_{p,k}^t = {{\bf{h}}_{p,k}}s_{p,k}^t + {\bf{w}}_{p}^t,
\end{array}
\end{equation}
in which ${{\bf{h}}_{p,k}}{ \in \mathbb{C}^{M \times 1}}$ is the subchannel associated with the $k$-th device at the $p$-th subcarrier, $s_{p,k}^t$, generated from i.i.d. standard complex Gaussian distribution, is the random access pilot from the $k$-th device, and ${\bf{w}}_{p}^t$  denotes the additive white Gaussian noise (AWGN) at the BS.
The channel vectors can be modeled as ${{\bf{h}}_{p,k}} = {\tau _k}{\widetilde {\bf{h}}_{p,k}}$, where ${\tau _{k}}$ is the large-scale fading caused by the path loss and shadowing fading, and ${\widetilde {\bf{h}}_{p,k}}$ is the small-scale fading channels \cite{{small-scale fading}}.
For the typical mMTC system, only a small subset of devices are activated to access the BS within a given coherent time interval, and the device activity indicator can be denoted as
\begin{equation}
{\alpha _{k}} = \left\{ \begin{array}{l}
\begin{array}{*{20}{c}}
{1,}&{{\rm{if \ the }}\ k\rm{th} {\rm{\ device\ is \ active}}},
\end{array}\\
\begin{array}{*{20}{c}}
{0,}&{{\rm{otherwise}}},
\end{array}
\end{array} \right.{\rm{   }}\forall k.
\end{equation}
Meanwhile, we define the set of active devices as ${\cal K} = \left\{ {k:{\alpha _k} = 1,\ k = 1, \cdots ,K} \right\}$ and the number of active devices is ${K_a}{\rm{ = }}\left| {\cal K} \right|$.
Hence, the received signals from all devices can be written as
\begin{equation}
\begin{array}{l}
{\bf{y}}_p^t = \sum\limits_{k = 1}^K {\alpha_k{{\bf{h}}_{p,k}}s_{p,k}^t + {\bf{w}}_p^t}  = {{\bf{H}}_p}{\bf{s}}_p^t + {\bf{w}}_p^t,
\end{array}
\end{equation}
in which ${{\bf{H}}_p} = [{\alpha _1}{{\bf{h}}_{p,1}}, \cdots ,{\alpha _K}{{\bf{h}}_{p,K}}]{ \in \mathbb{C}^{M \times K}}$, ${\bf{s}}_p^t = \left[ {s_{p,1}^t, \cdots ,s_{p,K}^t} \right]^{\rm{T}}{ \in \mathbb{C}^{K \times 1}}$.
Furthermore, the received signals over $G$ $(G \ll K)$ successive time slots are jointly exploited to detect the uncoordinated random access devices as
\begin{equation}\label{eq:cs problem}
\begin{array}{l}
{{\bf{Y}}_p} = {{\bf{S}}_p}{{\bf{X}}_p} + {{\bf{W}}_p},
\end{array}
\end{equation}
where ${{\bf{Y}}_p} = {\left[ {{\bf{y}}_p^1, \cdots ,{\bf{y}}_p^G} \right]^{\rm{T}}}{ \in \mathbb{C}^{G \times M}}$, ${{\bf{S}}_p} = {\left[ {{\bf{s}}_p^1, \cdots ,{\bf{s}}_p^G} \right]^T}{ \in \mathbb{C}^{G \times K}}$, ${{\bf{X}}_p} = {[{{\bf{x}}_{p,1}}, \cdots ,{{\bf{x}}_{p,K}}]^T}{ \in \mathbb{C}^{K \times M}}$, and ${{\bf{x}}_{p,k}} = {\alpha _k}{{\bf{h}}_{p,k}}$  denotes the channel vector of the $k$-th device at the $p$-th subcarrier.

In this paper, we consider the grant-free random access, where the indexes set ${\cal K}$ and channel vector ${{\bf{h}}_{p,k}}$ associated with the active devices can be jointly estimated from the noisy measurements ${{\bf{Y}}_p}$ and the random access pilot matrix ${{\bf{S}}_p}$ known at the BS.
On the other hand, due to the sporadic traffic of devices in typical mMTC, only a smaller number of devices are active, i.e., ${K_a} \ll K$.
This observation motivates us to formulate the joint active devices detection and channel estimation for massive random access as a CS problem.
Furthermore, we observe that different columns of ${{\bf{X}}_p}$ share the common support, and $\left\{ {{{\bf{X}}_p}} \right\}_{p = 1}^P$ have the common sparsity pattern, namely,
\begin{equation}\label{eq:common supp}
\begin{array}{l}
{\mathop{\rm supp}\nolimits} \left\{ {{{\bf{X}}_1}} \right\} = {\mathop{\rm supp}\nolimits} \left\{ {{{\bf{X}}_2}} \right\} = \cdots = {\mathop{\rm supp}\nolimits} \left\{ {{{\bf{X}}_P}} \right\},
\end{array}
\end{equation}
which inspires us to solve (\ref{eq:cs problem}) with the distributed multiple measurement vector (DMMV) CS theory for further improved performance \cite{{frequency common supp}}.

\section{CS-Based Massive Random Access Scheme}

In this section, we propose the DMMV-AMP algorithm for massive random access, where the sparse traffic observed from multiple antennas and multiple subcarriers is considered.
Especially, we integrate the hyperparameters learning manner of AMP with nearest neighbor sparsity pattern learning (AMP-NNSPL) algorithm \cite{{AMP_NNSPL_1}, {AMP_NNSPL_2}} into our scheme for further improved support estimation performance.

\subsection{DMMV-AMP Algorithm}

We first consider the massive random access problem for the $p$-th subcarrier, where the solution is listed in \textbf{Algorithm \ref{alg:AMP-NNSPL}} named MMV-AMP algorithm.
According to the theory of statistical signal processing, the minimum mean square error (MMSE) estimation of (\ref{eq:cs problem}) is the posterior means, which can be expressed as
\begin{equation}\label{eq:posterior mean 1}
\begin{array}{l}
{\hat x_{km}} = \int {{x_{km}}p({x_{km}}|\bf{Y})} d{x_{km}}, \ \forall k,m,
\end{array}
\end{equation}
where the subcarrier index $p$ in ${x_{p,km}}$ and ${\bf{Y}}_p$ is dropped to simplify the notations, and ${x_{p,km}}$ is the $\left( {k,m} \right)$-th element of the matrix ${{\bf{X}}_p}$.
The marginal posterior probability is calculated by $p({x_{km}}{\rm{|}}{\bf{Y}}) = \int {p({\bf{X}}|{\bf{Y}})} {d_{{{\bf{X}}_{\backslash km}}}}$, which involves the multi-dimensional integrals, and the ${{\bf{X}}_{\backslash km}}$ denotes the collection of $\left\{ {{x_{ij}}} \right\}_{1 \le j \le M,j \ne m}^{1 \le i \le K,i \ne k}$.
The joint posterior probability can be computed according to Bayesian rule in a factored form as
\begin{equation}
p({\bf{X}}|{\bf{Y}}) = \frac{1}{Z}\prod\limits_{g = 1}^G {\prod\limits_{m = 1}^M {p({y_{gm}}|{\bf{X}})} \prod\limits_{k = 1}^K {\prod\limits_{m = 1}^M {{p_0}({x_{km}})} } },
\end{equation}
in which $Z$ is the normalization factor.
In this paper, we consider a flexible spike and slab prior distribution which can match the real distribution of channels well
\begin{equation}
{p_0}({\bf{X}}) = \prod\limits_{k = 1}^K {\prod\limits_{m = 1}^M {\left[ {(1 - {\lambda _{km}})\delta ({x_{km}}) + {\lambda _{km}}f({x_{km}})} \right]}},
\end{equation}
where ${\lambda _{km}} \in (0,1)$ is the sparse ratio, i.e., the probability of ${x_{km}}$ being nonzero, $\delta ({x_{km}})$ is the Dirac delta function, $f({x_{km}})$ is the distribution of the nonzero entries.
Under the assumption of AWGN, the likelihood function of ${y_{gm}}$ can be expressed as
\begin{equation}\label{eq:likelihood function}
\begin{array}{l}
p({y_{gm}}|{\bf{X}}) = \frac{1}{{\sqrt {2\pi {\sigma ^2}} }}\exp \left( - \frac{1}{{2{\sigma ^2}}}{\left| {{y_{gm}} - \sum\limits_k {{S_{gk}}{x_{km}}} } \right|^2} \right),
\end{array}
\end{equation}
where ${\sigma ^2}$ is the variance of noise.
As the marginal probability is hard to compute, we resort to the belief propagation (BP), which provides low-complexity heuristics for approximating $p({x_{km}}{\rm{|}}{\bf{Y}})$.
A key remark is that AMP decouples the matrix estimation problem (\ref{eq:cs problem}) into $KM$ scalar problems in the asymptotic regime, i.e., $K\rightarrow\infty$ while $\lambda $ is fixed, and the posterior distribution of ${x_{km}}$ is approximated as  \cite{{AMP_NNSPL},{AMP 1}}
\begin{equation}\label{eq:decoupled posterior 1}
\begin{split}
p{\rm{(}}{x_{km}}|R_{_{km}}^t,\Sigma _{_{km}}^t{\rm{)  = }}\frac{1}{Z}{p_0}{\rm{(}}{x_{km}}{\rm{)}}{\cal C}{\cal N}{\rm{(}}{x_{km}};R_{_{km}}^t,\Sigma _{_{km}}^t{\rm{)}},
\end{split}
\end{equation}
where $t$ denotes the $t$-th iteration, $R_{_{km}}^t$ and $\Sigma _{_{km}}^t$ are calculated as step \ref{code:fram:variable nodes} in \textbf{Algorithm \ref{alg:AMP-NNSPL}}.
We have the assumption that $f({x_{km}}) = {\cal C}{\cal N}({x_{km}};\mu ,\tau )$, which is a flexible prior model for the channels.
By exploiting this prior model, the posterior distributions are obtained by (\ref{eq:decoupled posterior 1}) as
\begin{equation}\label{eq:decoupled posterior 2}
\begin{split}
p{\rm{(}}{x_{km}}|R_{_{km}}^t,\Sigma _{_{km}}^t{\rm{)}} &= {\rm{(1 - }}\pi _{_{km,}}^t{\rm{)}}\delta {\rm{(}}{x_{km}}{\rm{)}} \\                                            & + \pi _{_{km}}^t{\cal N}{\rm{(}}{x_{km}};{A_{_{km}}^t},{\Delta_{_{km}}^t}{\rm{)}},
\end{split}
\end{equation}
where
\[ {A_{_{km}}^t} = \frac{{\tau R_{km}^t + \Sigma _{km}^t\mu }}{{\Sigma _{km}^t + \tau }}, \ {\Delta_{_{km}}^t} = \frac{{\tau \Sigma _{km}^t}}{{\tau  + \Sigma _{km}^t}}, \]
\[ {\cal L} = \frac{1}{2}\ln \frac{{\Sigma _{km}^t}}{{\Sigma _{km}^t + \tau }} + \frac{{{{\left| {R_{km}^t} \right|}^2}}}{{2\Sigma _{km}^t}} - \frac{{{{\left| {R_{km}^t - \mu } \right|}^2}}}{{2(\Sigma _{km}^t + \tau )}}, \]
\begin{equation}\label{eq:equivalent sparse ratio}
\begin{array}{l}
\pi _{km}^t = \frac{{\lambda _{km}^t}}{{\lambda _{km}^t + (1 - \lambda _{km}^t)\exp ( - {\cal L})}},
\end{array}
\end{equation}
in which $\pi _{km}^t$ is the equivalent sparse ratio.
The posterior mean and posterior variance can now be explicitly calculated as
\begin{equation}
\begin{array}{l}
{g_a}\left( {R_{_{km}}^t,\Sigma _{_{km}}^t} \right) = \pi _{_{km}}^t{A_{_{km}}^t},
\end{array}
\end{equation}
\begin{equation}
\begin{array}{l}
{g_c}\left( {R_{_{km}}^t,\Sigma _{_{km}}^t} \right) = \pi _{_{km}}^t({\left| {{A_{_{km}}^t}} \right|^2} + {\Delta_{_{km}}^t}) - {\left| {{g_a}} \right|^2}.
\end{array}
\end{equation}
Thus, the AMP iterations can be calculated in the matrix-vector forms instead of the integrals.
However, conventional AMP algorithm assumes full knowledge of the prior distribution and noise variance, which is an impractical assumption.
Resorting to the EM algorithm, the DMMV-AMP algorithm can learn the unknown hyperparameters, i.e., $\bm{\theta} = \left\{ {\mu ,\tau ,{\sigma ^2},{\lambda _{km}},{\forall k,m}} \right\}$, which consists of two steps
\begin{equation}\label{eq:EM}
\begin{array}{l}
Q({\bm{\theta ,}}{{\bm{\theta }}^t}) = \mathbb{E}\left( {\ln p({\bf{X}}|{\bf{Y}})|{\bf{Y}};{{\bm{\theta }}^t}} \right),
\end{array}
\end{equation}
\begin{equation}
\begin{array}{l}
{{\bm{\theta }}^{t + 1}} = \arg \mathop {\max }\limits_{\bm{\theta }} Q\left({\bm{\theta ,}}{{\bm{\theta }}^t}\right),
\end{array}
\end{equation}
where $\mathbb{E}\left( { \cdot |{{\bf{Y}}};{{\bm{\theta }}^t}} \right)$ denotes expectation conditioned on measurements ${\bf{Y}}$ with parameters ${{\bm{\theta }}^t}$.
Two problems arise in the EM algorithm, the computation of $p({\bf{X}}|{\bf{Y}};{\bm{\theta }}^t)$ is high complexity and the joint optimization of ${\bm{\theta }}$ is difficult.
However, the approximation of $p({\bf{X}}|{\bf{Y}};{\bm{\theta }}^t)$ is given by the AMP as (\ref{eq:decoupled posterior 1}).
Moreover, we resort to incremental EM algorithm \cite{{Incremental EM}}, i.e., ${\bm{\theta }}$ is updated one element at a time while other parameters are fixed.
By taking the derivative of (\ref{eq:EM}) where the fixed parameters are considered as constant and zeroing the derivative, the update rules of the hyperparameters are obtained as
\begin{equation}\label{eq:Update noise variance}
\begin{array}{l}
{({\sigma ^2})^{t + 1}} = \frac{1}{{GM}}\sum\limits_g {\sum\limits_m {\left[ {\frac{{{{\left| {{y_{gm}} - Z_{gm}^t} \right|}^2}}}{{{{\left| {1 + V_{gm}^t/{{({\sigma ^2})}^t}} \right|}^2}}} + \frac{{{{({\sigma ^2})}^t}V_{gm}^t}}{{{{({\sigma ^2})}^t} + V_{gm}^t}}} \right]} },
\end{array}
\end{equation}

\begin{equation}\label{eq:Update prior variance}
\begin{array}{l}
{\tau ^{t + 1}} = \frac{{\sum\limits_k {\sum\limits_m {\pi _{km}^t\left[ {{{\left| {{\mu ^t} - {A_{_{km}}^t}} \right|}^2} + {\Delta_{_{km}}^t}} \right]} } }}{{\sum\limits_k {\sum\limits_m {\pi _{km}^t} } }},
\end{array}
\end{equation}

\begin{equation}\label{eq:Update prior mean}
\begin{array}{l}
{\mu ^{t + 1}} = \frac{{\sum\limits_k {\sum\limits_m {\pi _{km}^t{A_{_{km}}^t}} } }}{{\sum\limits_k {\sum\limits_m {\pi _{km}^t} } }},\  \lambda _{km}^{t + 1} = \pi _{km}^t.
\end{array}
\end{equation}
As the EM algorithm may converge to the local extremum of the likelihood function, the initialization of the hyperparameters is very important, which are given as
\begin{equation}\label{eq:Initial lamda}
\begin{array}{l}
\lambda _{km}^0 = {\lambda ^0} = \frac{G}{K}\left\{ {\mathop {\max }\limits_{c > 0} \frac{{1 - 2K\left[ {{{(1 + c)}^2}\Phi ( - c) - c\phi (c)} \right]/G}}{{1 + {c^2} - 2\left[ {{{(1 + c)}^2}\Phi ( - c) - c\phi (c)} \right]}}} \right\},
\end{array}
\end{equation}

\begin{equation}
\begin{array}{l}
{({\sigma ^2})^0} = \frac{1}{M}\sum\limits_m {\frac{{{{\left\| {{{\bf{y}}_m}} \right\|}^2}}}{{({\rm{SNR}^0} + 1)G}}},
\end{array}
\end{equation}

\begin{equation}\label{eq:Initial prior para}
\begin{array}{l}
{\tau ^0} = \frac{1}{M}\sum\limits_m {\frac{{\left\| {{{\bf{y}}_m}} \right\|_2^2 - M{{({\sigma ^2})}^0}}}{{\left\| {\bf{S}} \right\|_F^2{\lambda ^0}}}},\ {\mu ^0} = 0,
\end{array}
\end{equation}
where $\Phi ( - c)$ and $\phi (c)$ are, respectively, the cumulative distribution function and the probability density function of the standard normal distribution, $c$ denotes the maximum sparse ratio with fixed $G/K$, and $\rm{SNR^0} = 100$.

\begin{algorithm}[t]
\caption{MMV-AMP Algorithm}
\label{alg:AMP-NNSPL}
\begin{algorithmic}[1]
\REQUIRE Noisy observation ${\bf{Y}}_p$, pilot matrix ${\bf{S}}_p$, the maximum number of iteration ${T_{\rm{max}}}$ and termination threshold $\varepsilon $.
\ENSURE The estimated channel vectors ${\widehat {\bf{x}}_k}$, $\forall k$.
\STATE Set $t = 1$, $\hat x_{km}^1 = \int {{x_{km}}{p_0}({x_{km}})d{x_{km}}} $, $v_{km}^1 = \int {{{\left| {{x_{km}} - \hat x_{km}^1} \right|}^2}{p_0}({x_{km}})d{x_{km}}} $, $V_{gm}^0 = 1$, $Z_{gm}^0 = {y_{gm}}$, $\forall k,m,g$.  Initialize the hyperparameters as (\ref{eq:Initial lamda})-(\ref{eq:Initial prior para}).
\label{ code:fram:Initialization}
\REPEAT
\label{code:fram:repeat}
\STATE Factor nodes update: $g = 1, \cdots ,G;\ m = 1, \cdots ,M$
\\ $V_{gm}^t = \sum\limits_k {{{\left| {{{S}_{gk}}} \right|}^2}v_{km}^t}$
\\ $Z_{gm}^t = \sum\limits_k {{S_{gk}}\hat x_{km}^t - \frac{{V_{gm}^t}}{{{{\left( {{\sigma ^2}} \right)}^t} + V_{gm}^{t - 1}}}({y_{gm}} - Z_{gm}^{t - 1})}$
\label{code:fram:factor nodes}
\STATE Variable nodes update: $k = 1, \cdots ,K;\ m = 1, \cdots ,M$
\\ $\Sigma _{km}^t = {\left[ {\sum\limits_g {\frac{{{{\left| {{S_{gk}}} \right|}^2}}}{{{{\left( {{\sigma ^2}} \right)}^t} + V_{gm}^t}}} } \right]^{ - 1}}$
\\ $R_{km}^t = \hat x_{km}^t + \Sigma _{km}^t\sum\limits_g {\frac{{S_{gk}^*({y_{gm}} - Z_{gm}^t)}}{{{{\left( {{\sigma ^2}} \right)}^t} + V_{gm}^t}}}$
\\ $\hat x_{km}^{t + 1} = {g_a}(R_{km}^t,\Sigma _{km}^t)$
\\ $v_{km}^{t + 1} = {g_c}(R_{km}^t,\Sigma _{km}^t)$
\label{code:fram:variable nodes}
\STATE Update the hyperparameters as (\ref{eq:Update noise variance})-(\ref{eq:Update prior mean}), $t=t+1$.
\label{code:fram:hyperparameters update}

\UNTIL{$t > T_{\rm{max}}}$ or ${\left\| {{{\widehat {\bf{X}}_p}^{t + 1}} - {{\widehat {\bf{X}}_p}^t}} \right\|_{\rm{F}}} < \varepsilon {\left\| {{{\widehat {\bf{X}}_p}^t}} \right\|_{\rm{F}}}$.
\RETURN ${\hat x_{km}}$, $\forall k,m$
\end{algorithmic}
\end{algorithm}

Moreover, the MMV-AMP algorithm is further extended to the distributed model with all subchannel matrix ${{\bf{X}}_p}$, $\forall p \in \left[ P \right]$ being jointly estimated, which is termed as the DMMV-AMP algorithm.
For all subcarriers, the messages are parallelly updated as step \ref{code:fram:factor nodes}, step \ref{code:fram:variable nodes} in \textbf{Algorithm \ref{alg:AMP-NNSPL}}.
Notice that the columns of channel matrix ${{\bf{X}}_p}$ share the common support, which can be utilized to improve the accuracy of the support estimation.
We further exploit the common sparsity described in (\ref{eq:common supp}), and the estimated sparse ratio is updated by $\hat \lambda _{km}^{t + 1}{\rm{ = }}\hat \lambda _k^{t + 1} = \frac{1}{{MP}}\sum\limits_p {\sum\limits_m {\pi _{p,km}^t} } $.
The modified update rules of the mean and the variance of nonzero entries are written as
\begin{equation}\label{eq:mean and variance of mul}
\begin{array}{l}
{\mu ^{t + 1}} = \frac{1}{P}\sum\limits_p {\mu _p^t},\ {\tau ^{t + 1}} = \frac{1}{P}\sum\limits_p {\tau _p^t}.
\end{array}
\vspace*{-1.5mm}
\end{equation}

Additionally, compared with conventional AMP-based solutions, we consider the more practical frequency-selective fading channels, thus the equivalent signal-to-noise ratio (SNR) varies with the shubcarrier index $p$.
We assume there is an adaptive power control at the BS to eliminate the effect of the large scale fading, and all the subcarriers have the same signal power, thus the frequency-selective fading effect is included in the different noise variances $\left( {{\sigma ^2}} \right)_p^t$ which are respectively calculated as (\ref{eq:Update noise variance}).
With the estimated channel matrix ${\widehat {\bf{X}}_n}$ and the characteristic of the sparsity pattern, we develop a threshold-based activity detector defined as follows
\begin{equation}\label{eq:threshold detector of mul}
\begin{array}{l}
{{\hat \alpha _k}} = \left\{ {\begin{array}{*{20}{c}}
{1,}&{\sum\limits_p {\sum\limits_m {r(\hat x_{p,km}) \ge {p_{\rm{th}}}MP} } },\\
{0,}&{\sum\limits_p {\sum\limits_m {r(\hat x_{p,km}) < {p_{\rm{th}}}MP} } }.
\end{array}} \right.
\end{array}
\end{equation}
In (\ref{eq:threshold detector of mul}), $r( \cdot )$ denotes a mapper where $r({\hat x_{km}}) = 1$  if $\left| {{{\hat x}_{km}}} \right| > {10^{ - 10}}$, otherwise $r({\hat x_{km}}) = 0$, and $p_{\rm{th}}$ is a tunable parameter, which will influence the probability of missed detection and false alarm.
In this paper, we consider ${p_{\rm{th}}} = 0.99$.
Finally, given that the $k$-th device is declared active, its channel is estimated as ${\hat {\bf{h}}_{p,k}} = {\hat {\bf{x}}_{p,k}}$, where ${\hat {\bf{x}}_{p,k}}$ is the $k$-th row of the estimated channel matrix ${\widehat {\bf{X}}_p}$.

\subsection{State Evolution}
Statistical properties of AMP algorithms allow us to accurately analyze their performance in the asymptotic regime \cite{{AMP 2}}.
In this subsection, we use SE to characterize the mean square error (MSE) performance of the proposed scheme.
The MSE of the estimation and the variance of the estimated signal are defined as
\[ {E^t} = \frac{1}{{KM}}\sum\limits_k {\sum\limits_m {{{\left| {\hat x_{km}^t - x_{km}^t} \right|}^2}} },\ {V^t} = \frac{1}{{KM}}\sum\limits_k {\sum\limits_m {v_{km}^t} }. \]
Define the random variable ${X_0} \sim {p_0}\left( {\bf{X}} \right)$, ${\rm{Z}} \sim {\cal C}{\cal N}(z;0,1)$, then the mean and variance of the posterior distribution are respectively expressed as \cite{{AMP 2}}
\begin{equation}
 {R^t} = {x_0} + \sqrt {\frac{{\sigma _0^2 + {E^t}}}{{G/K}}} z,\ {\Sigma ^t} = \frac{{{{({\sigma ^2})}^t} + {V^t}}}{{G/K}},
\end{equation}
which show that the AMP algorithm decouples the vector or matrix estimation problem into independent scalar problems, $\sigma _0^2 = 1/\rm{SNR}$, and the ${E^t}$ and ${V^t}$ can be updated as
\begin{equation}
{E^t} = \mathbb{E}\left[ {{{\left| {{g_a}({R^t},\ {\Sigma ^t}) - {x_0}} \right|}^2}} \right],\ {V^t} = \mathbb{E}\left[ {{g_c}({R^t},{\Sigma ^t})} \right],
\end{equation}
with $\mathbb{E}\left[  \cdot  \right]$ denoting the expectation with respect to the random variable ${X_0}$ and $Z$.
In contrast to the conventional AMP algorithms assuming full knowledge of the prior distribution, the SE of DMMV-AMP agorithm also need to track the update rules of the hyperparameters ${\bm{\theta }}$, which are given as follows
\begin{equation}
{\pi ^{t + 1}} = \mathbb{E}\left( {{\pi ^t}} \right),\ {\tau ^{t + 1}} = \mathbb{E}\left( {{\tau ^t}} \right),\ {\mu ^{t + 1}} = \mathbb{E}\left( {{\mu ^t}} \right),
\end{equation}
\begin{equation}
{\left( {{\sigma ^2}} \right)^{t + 1}} = \frac{{\sigma _0^2 + {E^t}}}{{\left[ {{\rm{1 + }}{V^{t}}/{{\left( {{\sigma ^2}} \right)}^t}} \right]}} + \frac{{{{\left( {{\sigma ^2}} \right)}^t}{V^{t}}}}{{{{\left( {{\sigma ^2}} \right)}^t} + {V^{t}}}},
\end{equation}
where ${\pi ^t}$, ${\tau ^t}$, ${\mu ^t}$ are calculated as (\ref{eq:equivalent sparse ratio}), (\ref{eq:Update prior variance}) and (\ref{eq:Update prior mean}).

\section{Simulation Results}

In this section, we provide the simulated and analytical results of the proposed scheme.
Consider a uplink mMTC system with one BS equipped with $M$ = 32 antennas and $P$ pilots are uniformly allocated in $N$ = 2048 subcarriers.
We assume $K$ = 1000 potential devices are randomly distributed in the cell with radius 1km and ${K_a}$ = 100 devices are active at a time.
The carrier frequency is 2GHz, the bandwith is 10MHz, and the received signal-to-noise ratio is SNR = 20dB.
We set ${T_{\rm{max}}}$ = 200 and  $\varepsilon$ = ${10^{{\rm{ - }}5}}$.
The performance is evaluated by the probability of error detection ${P_e}$ and the normalized MSE (NMSE) under various time overhead $G$ and pilot sequence $P$, which are defined as
\[ {P_e} = \frac{{\sum\limits_k {\left| {{{\hat \alpha }_k} - {\alpha _k}} \right|} }}{K},\  {\rm{NMSE}} = 10\log \frac{{\sum\limits_p {\left\| {{{\widehat {\bf{X}}}_p} - {{\bf{X}}_p}} \right\|_{\rm{F}}^2} }}{{\sum\limits_p {\left\| {{{\bf{X}}_p}} \right\|_{\rm{F}}^2} }}. \]

\begin{figure}[t]
    \begin{minipage}[t]{0.45\columnwidth}
        \centering
        \includegraphics[width=1\columnwidth]{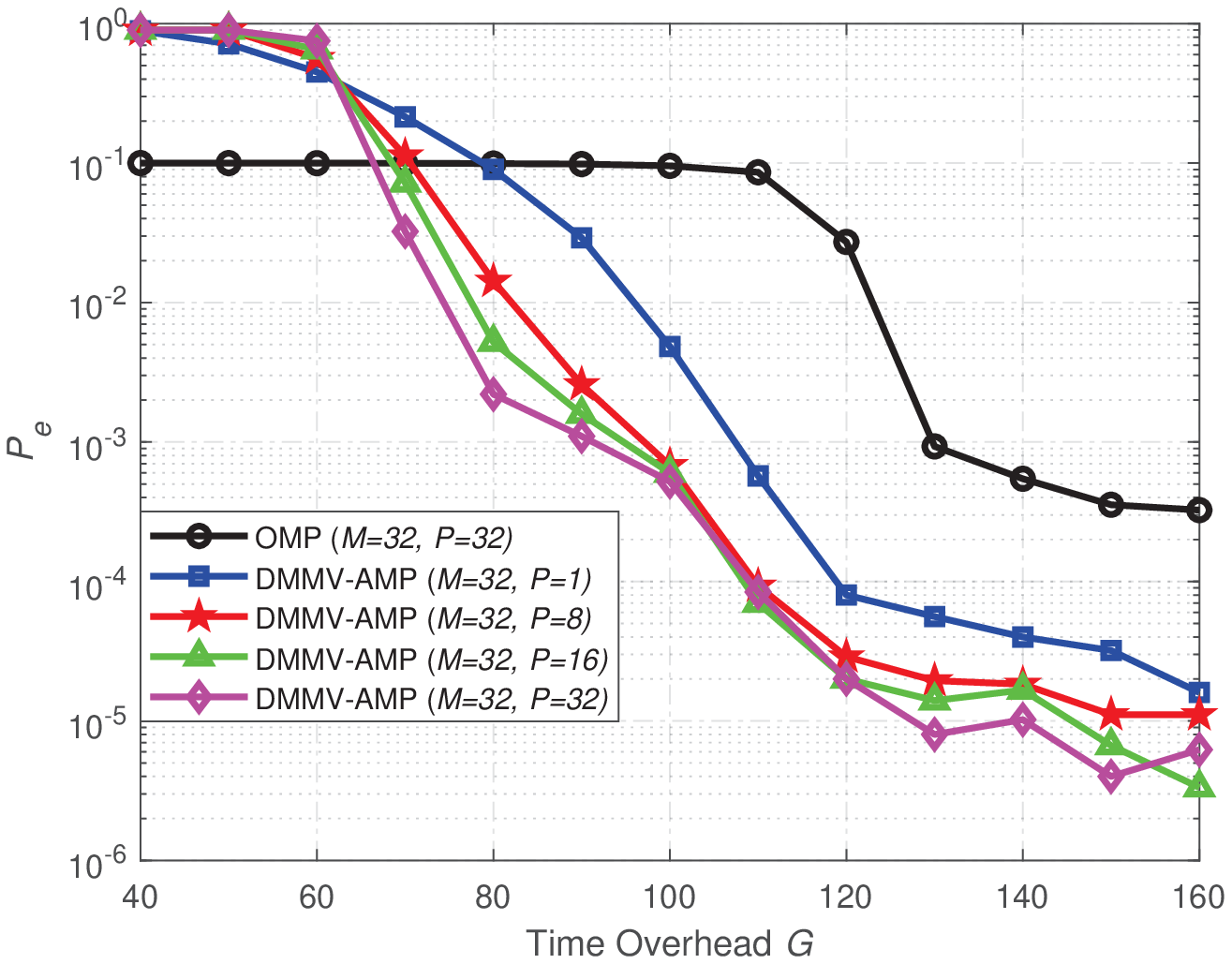}
        \vspace*{-5.5mm}
        \caption{The probability of error detection comparison as the function of $G$.} 
        \label{fig:ERR}
    \end{minipage}
    \hfill
    \begin{minipage}[t]{0.45\columnwidth}
        \centering
        \includegraphics[width=1\columnwidth]{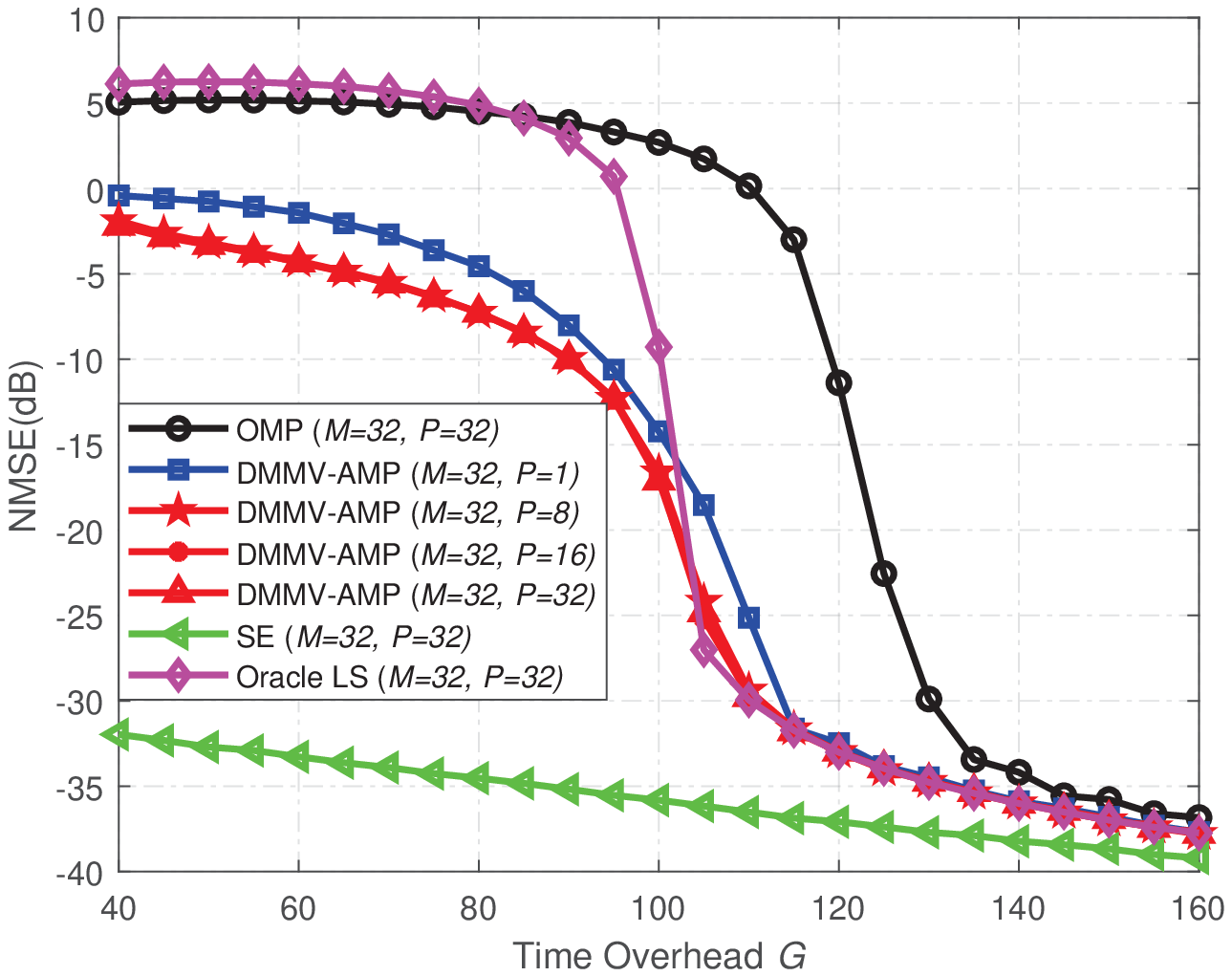}
        \vspace*{-5.5mm}
        \caption{The NMSE performance comparison and the SE of DMMV-AMP.} 
        \label{fig:NMSE}
    \end{minipage}
    \vspace*{-3.5mm}
\end{figure}

Fig. \ref{fig:ERR} examines the performance of device activity detector achieved by the DMMV-AMP algorithm and OMP algorithm \cite{{OMP}}.
It can be observed that the ${P_e}$ of DMMV-AMP based scheme decreases over $G$ rapidly, but ${P_e}$ of the OMP based scheme remains unchanged when $G<110$.
There exists a significant performance gap between the DMMV-AMP based scheme and the OMP based scheme when $G>80$, which shows that DMMV-AMP based scheme can significantly reduce the access latency when the same performance is considered.
Further, with the fixed $M$, when the common support among multiple carriers is leveraged, the performance can be further improved.

Fig. \ref{fig:NMSE} verifies the NMSE performance of the proposed scheme, OMP based scheme and the oracle LS based scheme with the known support set of the sparse channel matrix.
It shows that when time overhead is large enough, both DMMV-AMP algorithm and OMP algorithm can approach the performance of the oracle LS based scheme, since the support is estimated exactly in this case.
However, the proposed scheme outperforms the OMP based scheme when $G<140$, and its performance becomes better when $P$ increases.
An important observation is that when the pilot length is less then the active devices ($G<K_a$), the proposed scheme can still work very well by exploiting the structured sparsity shared by different subcarriers and receiver antennas.
The proposed DMMV-AMP based scheme can even outperforms the orale LS when $G<100$.
In addition, the performance of the proposed scheme is well predicted by the sate evolution when the time overhead is large enough.

\section{Conclusion}

A CS-based massive random access scheme has been proposed for uplink mMTC systems, which can significantly reduce the access latency.
By exploiting the structured sparsity among multiple BS antennas and multiple carriers, we propose a DMMV-AMP algorithm, and its SE is also derived to analyze the performance.
Simulation results demonstrate that the proposed scheme outperforms its counterparts with significantly reduced access latency.

\end{document}